\documentclass[prl,reprint,amsmath,amssymb]{revtex4-1}
\usepackage{graphicx}
\usepackage{dcolumn}
\usepackage{bm}
\usepackage{epsfig}
\usepackage{amssymb}
\usepackage{amsmath}
\newcommand{\Ro}{R_\mathrm{o}}

\begin{document}

\title{The probabilistic structure of the geodynamo}
\author{Christian R. Scullard}
\affiliation{Lawrence Livermore National Laboratory, Livermore, California, USA}
\author{Bruce A. Buffett}
\affiliation{Department of Earth and Planetary Science, University of California, Berkeley, California, USA}
\date{\today}

\begin{abstract}
One of the most intriguing features of Earth's axial magnetic dipole field, well-known from the geological record, is its occasional and unpredictable reversal of polarity. Understanding the phenomenon is rendered very difficult by the highly non-linear nature of the underlying magnetohydrodynamic problem. Numerical simulations of the liquid outer core, where regeneration occurs, are only able to model conditions that are far from Earth-like. On the analytical front, the situation is not much better; basic calculations, such as relating the average rate of reversals to various core parameters, have apparently been intractable. Here, we present a framework for solving such problems. Starting with the magnetic induction equation, we show that by considering its sources to be stochastic processes with fairly general properties, we can derive a differential equation for the joint probability distribution of the dominant toroidal and poloidal modes. This can be simplified to a Fokker-Planck equation and, with the help of an adiabatic approximation, reduced even further to an equation for the dipole amplitude alone. From these equations various quantities related to the magnetic field, including the average reversal rate, field strength, and time to complete a reversal, can be computed as functions of a small number of numerical parameters. These parameters in turn can be computed from physical considerations or constrained by paleomagnetic, numerical, and experimental data.
\end{abstract}

\maketitle

Understanding the generation and reversals of Earth's magnetic field is one of the enduring problems in geophysical and planetary science. Although it is clear enough by now that the source of the field is dynamo action in the liquid iron outer core, vigorous convection leads to highly unpredictable flow, making theoretical calculations of basic quantities, such as the average reversal rate, almost impossible. In the last two decades, some light has been shed by direct numerical simulations of Earth's core \cite{Glatzmaier1995,Kuang1999,Matsui2014} that feature a self-sustaining dipole-dominated field, that in many cases reverses at irregular intervals just like the real system. However, the predictive power of these calculations is severely hampered by the fact that they do not have sufficient resolution to use Earth-like parameters due to the core's very low viscosity. In these simulations, the Ekman number, a dimensionless measure of viscous effects, is orders of magnitude larger than what is thought to be its true value. This situation, barring some unforeseen revolution in computing power, is likely to persist for the foreseeable future.

An alternative approach is to model the geomagnetic field as a stochastic process, and there have been many models of this type over the years \cite{Hoyng2001,Petrelis2009,Mori2013,Buffett2014,Buffett2015,Buffett2017}. These are usually predicated on exploiting the qualitative similarity between paleomagnetic data and some well-understood or easily-studied stochastic process. We aim here to develop a stochastic differential equation that is derived directly from the underlying equations of the system. That is, we will not appeal to mean field theory, as in a previous approach of Hoyng et al. \cite{Hoyng2001}, or impose a Langevin form \cite{Buffett2014}; rather, our starting point will be the magnetic induction equation with a random source, which at heart is all the dynamo system really is. This stochastic equation was previously written down by Parker \cite{Parker1955} but to our knowledge its consequences have not been fully explored. We show that, when one considers only the slowest-decaying modes of the most important poloidal and toroidal components and makes fairly generic assumptions about the fluctuations, one can derive a differential equation for the probability distribution, a type of master equation. Furthermore, we show that assuming small-amplitude fluctuations leads to a Fokker-Planck equation in the toroidal and poloidal field amplitudes. This equation can then be used to compute various averages of interest in paleomagnetism, such as the time between reversals, the strength of the dipole field, and the duration of reversals, in terms of a small number of parameters.

The starting point of our analysis is the induction equation for the magnetic field in Earth's core,
\begin{equation}
 \frac{\partial {\bf B}}{\partial t}=\nabla \times ({\bf v} \times {\bf B})+\eta \nabla^2 {\bf B} \label{eq:induction}
\end{equation}
where $\eta = 1/(\sigma_e \mu_0)$, $\sigma_e$ is the electrical conductivity of the fluid outer core and $\mu_0$ is the permeability. The fluid velocity, ${\bf v}({\bf r},t)$, satisfies the Navier-Stokes equation with source terms accounting for temperature or concentration gradients as well as the Lorentz force resulting from the magnetic field. The difficulties associated with the geodynamo problem stem from this complicated interaction between the magnetic, velocity, and temperature or concentration fields.

We use the standard decomposition into poloidal and toroidal fields,
\begin{equation}
 {\bf B}={\bf B}_P+{\bf B}_T
\end{equation}
where
\begin{eqnarray}
 {\bf B}_P &=& \nabla \times \nabla \times P(r,\theta,\phi,t) {\bf \hat{r}} \\
 {\bf B}_T &=& \nabla \times  T(r,\theta,\phi,t) {\bf \hat{r}} .
\end{eqnarray}
The functions $P$ and $T$ are expanded in spherical harmonics,
\begin{eqnarray}
 P(r,\theta,\phi,t)&=&\sum_{l=0}^{\infty}\sum_{m=-l}^{l} b^P_{lm}(r,t) Y_{lm}(\theta,\phi) \\
 T(r,\theta,\phi,t)&=&\sum_{l=0}^{\infty}\sum_{m=-l}^{l} b^T_{lm}(r,t) Y_{lm}(\theta,\phi).
\end{eqnarray}
We will focus our attention on the poloidal dipole field, $l=1,m=0$, which is by far the dominant field seen at the surface, and a quadrupolar toroidal field from which the dipole is generated. The equations for these are found by inserting their expansions into the induction equation \cite{Kuang1999}, and after rescaling the length by $R_o$, the radius of the outer core, and the time by the diffusion time, $R_o^2/\eta$, we find
\begin{eqnarray}
 \frac{\partial}{\partial t}b^P_{10}(r,t) - \left[\frac{\partial^2}{\partial r^2}-\frac{2}{r^2} \right] b^P_{10}(r,t)&=&\frac{r^2}{2}f^P_{10}(r,t) \label{eq:b10P} \\
 \frac{\partial}{\partial t}b^T_{20}(r,t) - \left[\frac{\partial^2}{\partial r^2}-\frac{6}{r^2} \right] b^T_{20}(r,t)&=&\frac{r^2}{6}f^T_{20}(r,t) \label{eq:b20T}
\end{eqnarray}
The sources, $f^P_{10}(r,t)$ and $f^T_{20}(r,t)$, contain all the higher-order magnetic and velocity field harmonics. We treat these source terms as stochastic processes but, as we are interested in fluctuating, as well as average, properties we do not resort to mean field theory. Rather, a formal solution of the stochastic differential equation will be our starting point. To facilitate a further simplification, we expand the fields in spherical Bessel functions, the eigenfunctions of the induction operator,
\begin{eqnarray}
  b^P_{10}(r,t)=\sum_{n=1}^\infty c^P_n(t) \sigma_n r j_1 \left(\sigma_n r \right) \label{eq:bPexp} \\
  b^T_{20}(r,t)=\sum_{n=1}^\infty c^T_n(t) \mu_n r j_2 \left(\mu_n r \right) \label{eq:bTexp}  
\end{eqnarray}
where $j_n(r)$ is the $n^{\mathrm{th}}$ spherical Bessel function of the first kind and $\sigma_n$ and $\mu_n$ are constants that are determined from the boundary conditions. The condition that the poloidal field must be continuous with the irrotational field external to the outer core gives \cite{Parker1955,Elsasser1946}
\begin{equation}
 \sigma_n=n \pi,
\end{equation}
while the vanishing of the toroidal field at the core-mantle boundary leads to
\begin{equation}
 j_2(\mu_n)=0 .
\end{equation}
The first few $\mu_n =\{5.76,9.1,12.3...\}$. The decay rate of mode $n$ is $\sigma_n^2 \eta/R_o^2$ for the poloidal field, $\mu_n^2 \eta/R_o^2$ for the toroidal, and $n=1$ is by far the slowest-decaying mode for both. Their amplitudes satisfy the equations
\begin{eqnarray}
 \frac{d c_1^P}{dt}+\frac{\sigma_1^2}{\Ro^2}c_1^P(t)=\sum_{i=0}^{N_P} g_{i}^P \delta(t-t_i^P)+\Gamma_T(t) \label{eq:c1P} \\ 
 \frac{d c_1^T}{dt}+\frac{\mu_1^2}{\Ro^2}c_1^T(t)=\sum_{i=0}^{N_T} g_{i}^T \delta(t-t_i^T)+\Gamma_P(t) . \label{eq:c1T}
\end{eqnarray}
where $\Gamma_T(t)$ and $\Gamma_P(t)$ are noise terms, $t_i$ denotes the times of convective events that add to the respective component and $g_{i}$ their contribution. Here, we have assumed that these events occur on a time scale much shorter than those of interest in geomagnetism (hence the delta functions), such as the average time between reversals. The occurrence times are taken to be Poisson processes, to be described below, and the associated $g_i$ are also random variables. To model the generation of the two fields from each other, we assume the $g_i$ to be of the form
\begin{eqnarray}
 g_i^P&=&A_i^P c_1^T(t_i) f_T(c_1^T)  \label{eq:giP} \\
 g_i^T&=&A_i^T c_1^P(t_i) f_P(c_1^P). \label{eq:giT}
\end{eqnarray}
That is, to compute the source of the poloidal (toroidal) field, a convective event is associated with an amplitude $A_P$ ($A_T$) which we multiply against the present toroidal (poloidal) field. These events are essentially flows of non-zero helicity, and, although we will not consider any specific models of them, the amplitudes are related to the various properties of these flows such as their energy and angular momentum. Here, we will simply take the $A$ to be random variables. The functions $f_T$ and $f_P$ above are quenching functions; if the fields become large, Lorentz forces oppose regeneration and, for example, $g_i^P \rightarrow 0$ as $c_1^T \rightarrow \infty$. It is necessary to model this effect in some way for the system to have stable, non-zero magnetic fields. Again, we will not be too specific about this, and assume only that these are non-increasing functions of their arguments. The noise terms in (\ref{eq:c1P}) and (\ref{eq:c1T}) represent sources of fluctuations that do not contribute on average to the magnetic fields. As such, we assume
\begin{equation}
 \langle \Gamma_P(t) \rangle= \langle \Gamma_T(t) \rangle=0 .
\end{equation}
Although this noise has zero average, its fluctuations can be a significant influence on geomagnetic time scales, affecting quantities such as the average reversal rate and the time taken to complete a reversal. We shall make the assumption that the noise is a Gaussian process with correlation functions
\begin{eqnarray}
 \langle \Gamma_P(t) \Gamma_P(t') \rangle = q_P^2 \delta(t-t') \\
 \langle \Gamma_T(t) \Gamma_T(t') \rangle = q_T^2 \delta(t-t') .
\end{eqnarray}
The noise strengths $q_P$ and $q_T$ can in principle depend on the magnetic fields, but as evidence from both paleomagnetic \cite{Buffett2015} and numerical simulation data suggests the dependence is likely weak, we do not explicitly include this here.

Equations (\ref{eq:c1P}) and (\ref{eq:c1T}) can be formally solved; for example,
\begin{equation}
 c_1^P(t)=\sum_{i=0}^{N_P} g_{i}^P e^{-\sigma_1^2 (t-t_i^P)}.
\end{equation}
That is, the mode is usually decaying at the rate $\sigma_1^2$ but at the times $t_i^P$ the finite quantities $g_i^P$ are added to $c^P_1(t)$. This process has two stochastic components; the events occur at random times and they have random amplitudes. The arrival times, $t_i^P$ and $t_i^T$, are taken to be Poisson processes. Three of the defining features of such processes are \cite{Grimmett} 1) the probability that exactly one event occurs in an interval $\Delta t$ is $\xi \Delta t$ where $\xi$ is the rate (or intensity), 2) the probability that more than one event occurs is $\mathrm{O}(\Delta t^2)$, and 3) events occur independently of one another. The events will actually be taken to be an ensemble of Poisson processes, each member having its own amplitude $A$ and rate function $\xi(A)dA$. It will not be necessary have explicit forms for these, but one feature of the functions $\xi(A)$ can be immediately deduced; they are not symmetric under $A \rightarrow -A$ for the geodynamo. The Earth's magnetic field remains for long durations in the same polarity between apparently sudden reversals. This would seem to imply that regenerative events, characterized by positive $A$, are far more common than degenerative. As $\xi(A)dA$ is the rate of processes with amplitude between $A$ and $A+dA$, the total rates of all the different amplitudes are given by
\begin{eqnarray}
 Z^P \equiv \int_{-\infty}^\infty \xi^P(A)dA \\
 Z^T \equiv \int_{-\infty}^\infty \xi^T(A)dA
\end{eqnarray}
so, for example, $Z^P \Delta t$ is the probability that an event of any amplitude regenerates the poloidal field in the interval $\Delta t$.

With these preliminaries out of the way, we turn now to the time evolution of the probability for the process. The joint distribution, $P(x,y,t)dxdy$, is the probability at time $t$ that $c^P$ has a value between $x$ and $x+dx$, and  $c^T$ has a value between $y$ and $y+dy$. In the following analysis, we will set $\Gamma_T=\Gamma_P=0$. We assume these noise processes are uncorrelated with the fluctuations that regenerate the field, and thus their contribution can simply be added to the diffusion coefficient at the end of the derivation. Let us now think about what happens in this process during a small time interval $\Delta t$. With probability $1-(Z^P+Z^T) \Delta t$, there are no regeneration events in the interval. In this case, both fields simply decay at their natural rates so that we have
\begin{equation}
 P(x,y,t+\Delta t)dxdy=P(x',y',t)dx'dy'
\end{equation}
where
\begin{eqnarray}
 x'&\equiv& xe^{\sigma_1^2 \Delta t}\\
 y'&\equiv& ye^{\mu_1^2 \Delta t}.
\end{eqnarray}
The other possibility for the interval $\Delta t$ is the occurrence of either a poloidal or toroidal regeneration event, of amplitude between $A$ and $A+dA$. It is possible that more than one event occurs in $\Delta t$, but the probability of this is of order $(\Delta t)^2$ by the assumption of a Poisson process, so we may neglect it. The probability of a single poloidal source is $\xi^P(A)dA \Delta t$, and likewise $\xi^T(A)dA \Delta t$ for toroidal. Now, if an event of amplitude $A$ adds to the poloidal field we have
\begin{equation}
 P(x,y,t+\Delta t)dxdy=P(x-Ay f_T,y,t)dx dy ,
\end{equation}
that is, the probability that $c^P$ is at $x$ at time $t+\Delta t$ is simply the probability that it was at $x-Ayf_T(y)$, that is $x$ minus the quantity added by the event, at time $t$. A similar formula holds for the toroidal field. Putting everything together we have,
\begin{eqnarray}
& & P(x,y,t+\Delta t)dxdy= (1-Z^T \Delta t - Z^P \Delta t) \cr 
& & \ \ \ \ \times P\left(xe^{\sigma_1^2 \Delta t},ye^{\mu_1^2 \Delta t},t \right)e^{\sigma_1^2 \Delta t} e^{\mu_1^2 \Delta t}dxdy \cr
 & & \ \ \ \ + \int_{-\infty}^\infty \xi^P(A)P(x-Ay f_T(y),y,t)dA \Delta t dx dy \cr
 & & \ \ \ \ + \int_{-\infty}^\infty \xi^T(A)P(x,y-Ax f_P(x),t)dA \Delta t dx dy 
\end{eqnarray}
where we have integrated over all possible values of $A$ that can appear in the toroidal and poloidal amplitudes. Expanding everything to first order in $\Delta t$ and taking the limit $\Delta t \rightarrow 0$, we have the integro-differential equation,
\begin{eqnarray}
 \frac{\partial P}{\partial t}&=&\sigma_1^2 \frac{\partial}{\partial x}(xP)+\mu_1^2 \frac{\partial}{\partial y}(yP)-(Z^P+Z^T)P \cr
 &+&\int_{-\infty}^\infty \xi^P(A)P(x-Ay f_T(y),y,t)dA \cr 
 &+&\int_{-\infty}^\infty \xi^T(A)P(x,y-Ax f_P(x),t)dA . \label{eq:Pmaster}
\end{eqnarray}
This is a kind of master equation for the probability distribution. It is linear in $P$ and conserves normalization, as it must. Although it is surely possible to study this equation numerically, we will now discuss an important regime, namely the one in which the quantities added during an event, $A f$, are small. In this case, the distribution function satisfies the more familiar Fokker-Planck equation.

When $A f \ll 1$, the distribution can be expanded in $A$,
\begin{eqnarray}
 P(x-Ayf_T,y,t) &\approx& P(x,y,t)-Ay f_T(y) \frac{\partial P}{\partial x} \cr
 & & \ +\frac{A^2 y^2 f_T(y)^2}{2}\frac{\partial^2 P}{\partial x^2}
\end{eqnarray}
and similarly for $P(x,y-Axf_P,t)$. The result is the Fokker-Planck equation,
\begin{eqnarray}
 \frac{\partial P}{\partial t}&=&-\frac{\partial}{\partial x}[D^{(1)}_x(x,y)P]-\frac{\partial}{\partial y}[D^{(1)}_y(x,y)P] \cr
 & &+ D_x^{(2)}(y)\frac{\partial^2 P}{\partial x^2}+ D_y^{(2)}(x)\frac{\partial^2 P}{\partial y^2} \label{eq:FP}
\end{eqnarray}
where the drift coefficients are
\begin{eqnarray}
 D^{(1)}_x(x,y)&=&-\sigma_1^2 x+\langle A_P \rangle y f_T(y) \\
 D^{(1)}_y(x,y)&=&-\mu_1^2 y+\langle A_T \rangle x f_P(x)
\end{eqnarray}
and the diffusion coefficients, after we have added the parts arising from the additive terms $\Gamma_T(t)$ and $\Gamma_P(t)$,
\begin{eqnarray}
 D^{(2)}_x(y)&=&\frac{1}{2} \langle A_P^2 \rangle y^2 [f_T(y)]^2+\frac{1}{2}q_P^2 \\
 D^{(2)}_y(x)&=&\frac{1}{2} \langle A_T^2 \rangle x^2 [f_P(x)]^2+\frac{1}{2}q_T^2
\end{eqnarray}
with
\begin{eqnarray}
 \langle A_P \rangle &\equiv& \int_{-\infty}^\infty \xi^P(A)AdA\\
 \langle A_T \rangle &\equiv& \int_{-\infty}^\infty \xi^T(A)AdA\\
 \langle A_P^2 \rangle &\equiv& \int_{-\infty}^\infty \xi^P(A)A^2 dA\\
 \langle A_T^2 \rangle &\equiv& \int_{-\infty}^\infty \xi^T(A)A^2 dA .
\end{eqnarray}
Equation (\ref{eq:FP}) contains all the information about the process, and its solutions can be used to extract various properties, such as the rate of reversals, the average time of a reversal and the variation of these with core parameters. 

\begin{figure}
 \begin{center}
 \includegraphics[width=3.25in]{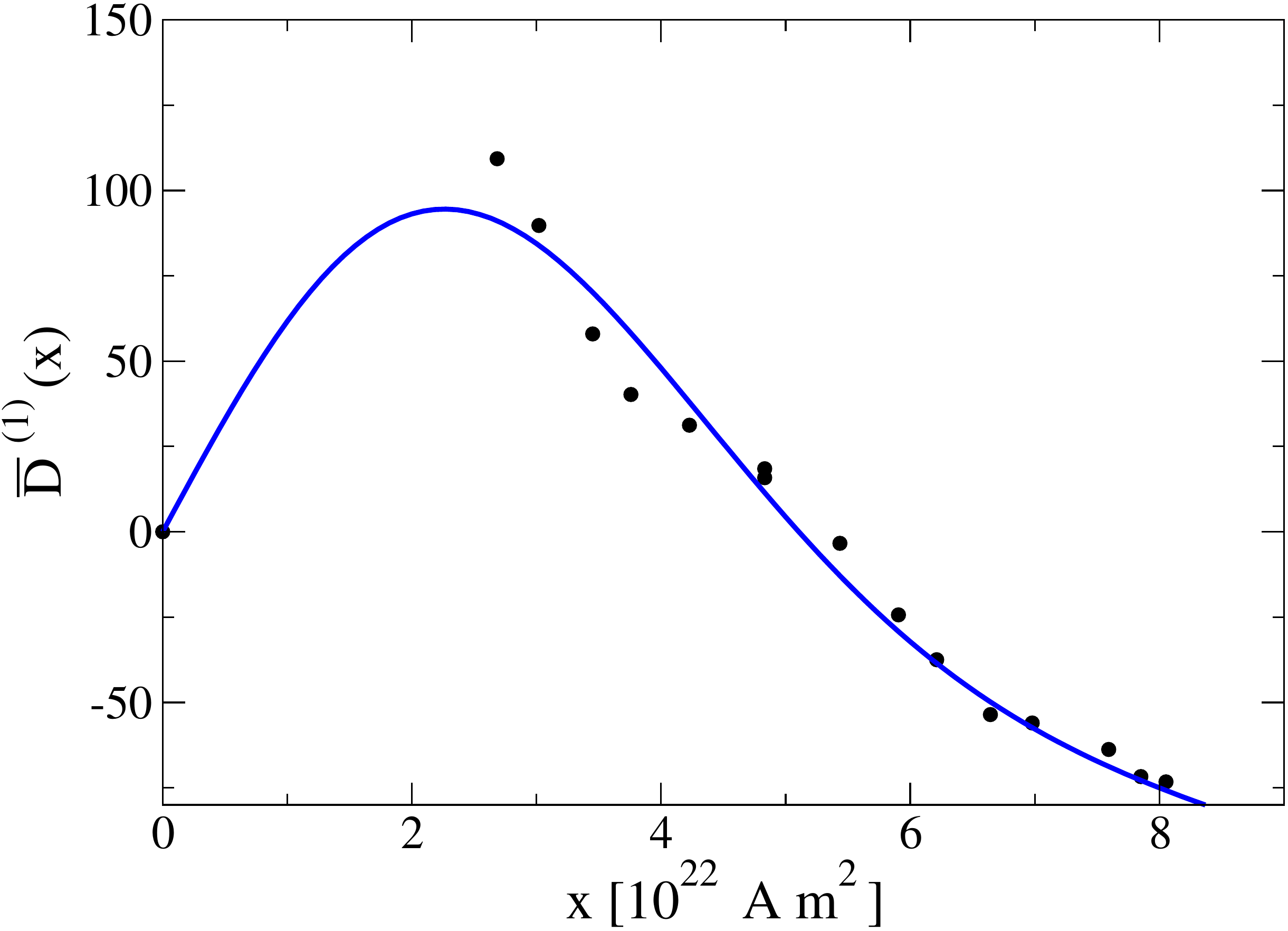}
 \end{center}
 \caption{The drift coefficient, $\overline{D}^{(1)}(x)$ in Eq. (\ref{eq:driftx}) fit to the PADM2M data set. Where the curve is positive, regeneration of the field is effective against its decay.}
 \label{fig:D1bar}
\end{figure}

Although $P(x,y,t)$ is the joint probability for the poloidal dipole and toroidal quadrupole fields, only the poloidal field is observed at Earth's surface. A natural question is now whether we can derive a Fokker-Planck equation, $P(x,t)$, for the dipole amplitude alone. In general, the answer is no; if we can imagine reversing the sign of only, say, the toroidal field, the statistics of the poloidal field would then change dramatically (we may, for example, have triggered a reversal of the poloidal field). However, given the somewhat disparate decay rates of the poloidal and toroidal modes, we can derive an approximate $P(x,t)$ using adiabatic elimination. The idea is that because the toroidal field decays the faster of the two, we can assume that it takes on a quasi-steady value that depends only on the current poloidal field; it is sometimes said that the fast variable is ``slaved'' to the slow one \cite{Risken}. The systematic determination of $P(x,t)$ from $P(x,y,t)$ under this approximation is straightfoward but somewhat tedious so we omit the details here. Our calculation is essentially identical to the one given in section 8.3 of Risken \cite{Risken}. Note that this procedure actually gives an expansion in $1/\lambda$ where $\lambda$ is the decay rate of the fast variable. Because the disparity between the poloidal and toroidal decay rates is only a factor of three, these $1/\lambda$ corrections might prove to be important. However, in what follows we keep only the leading-order term, neglecting those of order $1/\lambda$ and higher. To carry this calculation through, we assume the quenching functions have the form
\begin{eqnarray}
 f_T(x)&=&e^{-\gamma_T x^2} \\
 f_P(y)&=&e^{-\gamma_P y^2} .
\end{eqnarray}
The result is the Fokker-Planck equation for $P(x,t)$,
\begin{eqnarray}
 \frac{\partial P(x,t)}{\partial t}&=&-\frac{\partial}{\partial x} \left(\overline{D}^{(1)}(x) P(x,t) \right) \cr 
 & & \ \ \ \ \ +\frac{\partial^2}{\partial x^2}\left(\overline{D}^{(2)}(x) P(x,t) \right). \label{eq:FPx}
\end{eqnarray}
To get simple forms for the diffusion and drift coefficients, we set $\gamma_T=0$, and find
\begin{eqnarray}
\overline{D}^{(1)}(x)&=&-\sigma_1^2 x+\frac{\langle A_P \rangle \langle A_T \rangle}{\mu_1^2} x e^{-\gamma_P x^2}\label{eq:driftx} \\
\overline{D}^{(2)}(x)&=&\frac{\langle A_P^2 \rangle}{2 \mu_1^2}\left(\frac{\langle A_T^2 \rangle}{2}+\frac{\langle A_T \rangle^2}{\mu_1^2} \right) x^2 e^{-2 \gamma_P x^2} \cr
&+& \frac{1}{2}\left(\frac{q_T^2}{\mu_1^2}+q_P^2\right) .\label{eq:diffusionx}
\end{eqnarray}
The drift term (\ref{eq:driftx}) describes the generation of the dipole field. A typical example is shown in Figure \ref{fig:D1bar}. Where the curve is positive, the regeneration rate of the field is greater than its decay. The value of $x=x_0>0$ such that
\begin{equation}
 \overline{D}^{(1)}(x_0)=0 \label{eq:x0}
\end{equation}
is the typical value of the dipole field between reversals. Note that if $x_0$ is a solution of (\ref{eq:x0}) then so too is $-x_0$, consistent with the symmetry of the magnetohydrodynamic equations. It is possible that there is no positive region, and in this case dynamo action is not sufficient to support the dipole field and there is no $x_0>0$. This would occur whenever 
\begin{equation}
\left. \frac{d \overline{D}^{(1)}(x)}{dx} \right|_{x=0} < 0
\end{equation}
and from (\ref{eq:driftx}) we can therefore show that the dipole field is only regenerated when
\begin{equation}
 \langle A_P \rangle \langle A_T \rangle > \sigma_1^2 \mu_1^2 . \label{eq:critical}
\end{equation}

\begin{figure}
 \begin{center}
 \includegraphics[width=3.25in]{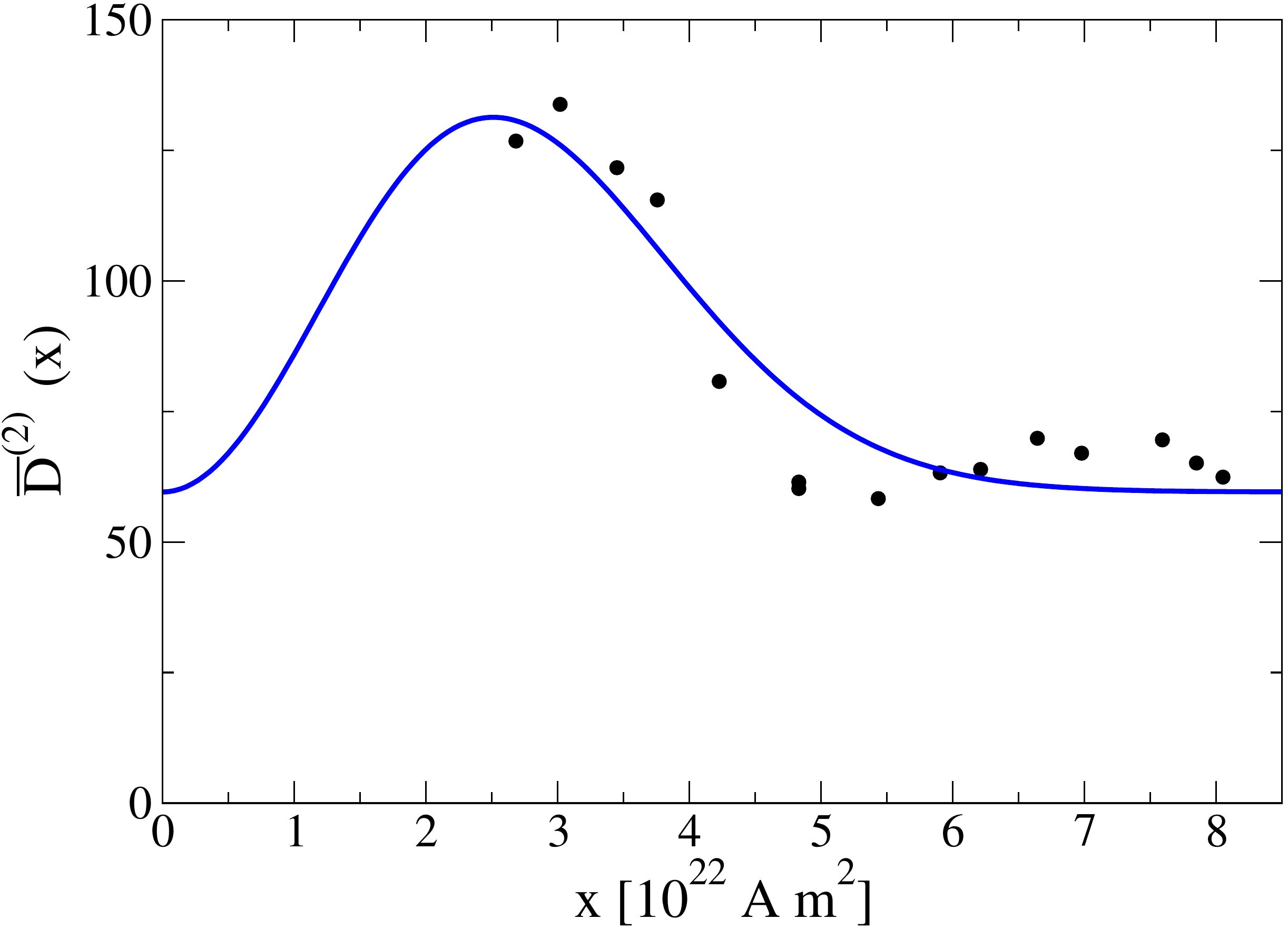}
 \end{center}
 \caption{The diffusion coefficient, $\overline{D}^{(2)}(x)$ in Eq. (\ref{eq:diffusionx}), fit to the PADM2M data set.}
 \label{fig:D2bar}
\end{figure}

\begin{figure}
 \begin{center}
 \includegraphics[width=3.0in]{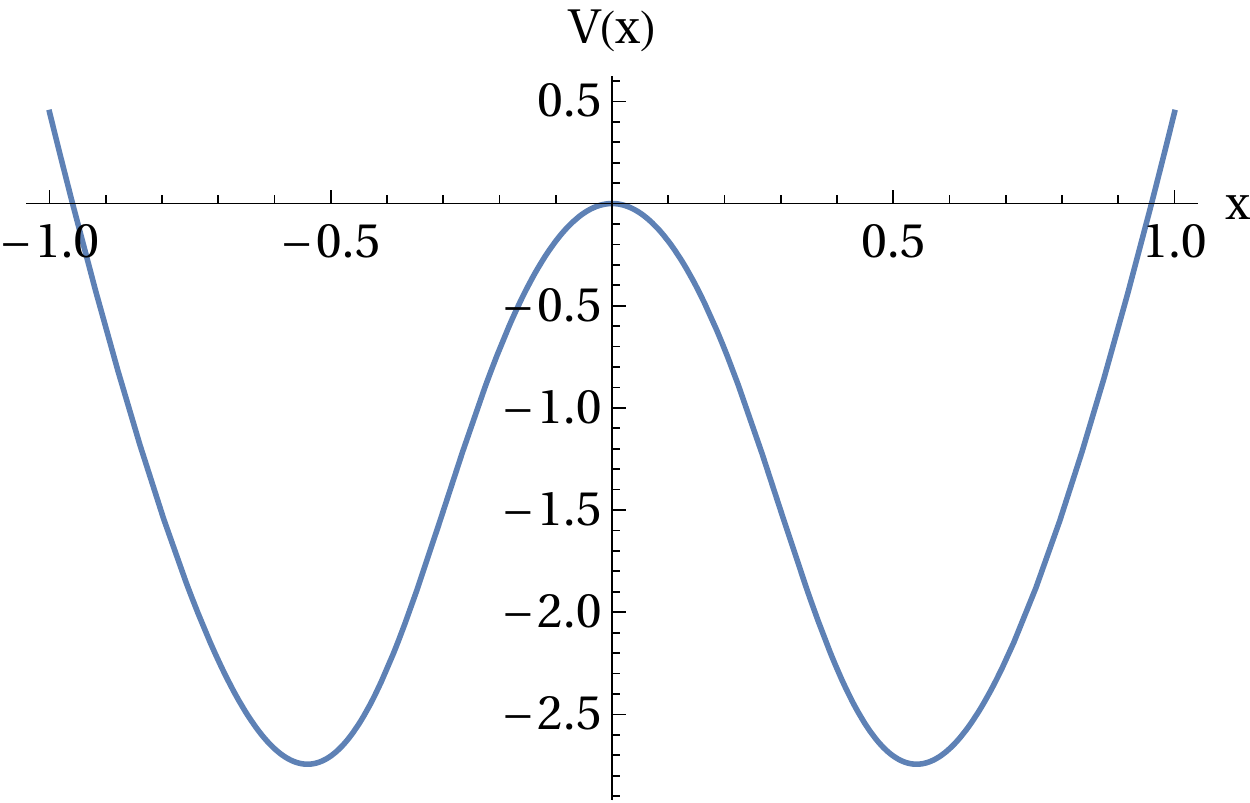}
 \end{center}
 \caption{Double-well potential derived from Fig. \ref{fig:D1bar} (after a change of variable described in the text).}
 \label{fig:potential}
\end{figure}

The formulas (\ref{eq:driftx}) and (\ref{eq:diffusionx}) for the drift and diffusion coefficients can be compared with data. In Fig. \ref{fig:D1bar}, we fit our Eq. (\ref{eq:driftx}) to the drift coefficient calculated by one of us \cite{Buffett2015} from the PADM2M model, a data set constructed from observations of Earth's dipole \cite{Ziegler}. The fit appears to be very good, and thus the formula we have derived is well supported by the data. We stress that $\overline{D}^{(1)}(x)$ must be odd in $x$ by the symmetry of the MHD equations, so if there was sufficient small-field data available it would look similar to our formula, at least qualitatively. In Fig. \ref{fig:D2bar} we plot the diffusion coefficient, where the fit uses the same parameters as those that produced Fig. \ref{fig:D1bar}. Once again, our formula seems to capture the main features of the PADM2M model. However, here it is not as clear what will happen in the small-field region where there is no data. The diffusion coefficient is even in $x$ and the shape of our curve arises from the assumption that the noise amplitudes $q_T$ and $q_P$ do not depend on the field. This is apparently a reasonable approximation for larger fields, but whether it holds as $x$ goes to zero is unclear.

Finally, we turn to the question of the reversal rate. For this, we can exploit the analogy between (\ref{eq:FPx}) and the equation for a heavily-damped particle in a potential, $V(x)$ \cite{Risken}, where $\overline{D}^{(1)}(x)=-\nabla V(x)$. A slight complication here is that in order to make contact with existing theory, we must transform from $x$ to a new variable in terms of which the diffusion coefficient is a constant, $D$. Details of this transformation are omitted but the procedure is laid out in Ref. \cite{Risken}. Choosing $D=1$, we find the potential plotted in Fig. \ref{fig:potential}. The Kramers escape formula gives the approximate reversal rate in the limit of a deep well \cite{Risken} in the sense that $\Delta V \equiv V(0)-V(x_{\mathrm{min}}) \gg D$,
\begin{equation}
 r=\frac{1}{2 \pi}\sqrt{|V''(0)|V''(x_{\mathrm{min}})}\exp(-\Delta V/D) . \label{eq:reversalrate}
\end{equation}

Using an electrical conductivity $\sigma_e = 1.2 \times 10^6\ \mathrm{\Omega} \cdot \mathrm{m}$ (needed to convert from dimensionless time), consistent with recent density functional theory calculations (DFT) \cite{Pozzo}, we find $r \approx 0.7\ \mathrm{Myr}^{-1}$. This $r$ is a bit low, but there are many areas of uncertainty here. For example, numerical \cite{Pozzo,Xu} and experimental \cite{Ohta,Konopkova} results have produced a range of electrical conductivities for the core, and the issue appears far from settled. Halving the conductivity doubles the computed reversal rate, although this is still lower than the current paleomagnetic estimates of $3$-$4\ \mathrm{Myr}^{-1}$ \cite{Gee,Lowrie}. A much greater source of uncertainty lies in the noise amplitudes, $q_T$ and $q_P$. As demonstrated in Ref. \cite{Buffett2017}, estimates of these kinds of quantities are highly error-prone and can vary widely between data sets. If, for example, $q_T$ and $q_P$ are doubled, then, using the conductivity from DFT, $r \approx 4\ \mathrm{Myr}^{-1}$. We can therefore obtain a realistic reversal rate by making plausible variations to the conductivity and noise, and perhaps a more sophisticated calculation will eventually be used to constrain those quantities.

Starting with the magnetohydrodynamic equations for Earth's core, we have derived a set of stochastic differential equations, characterized by a handful of parameters, governing the evolution of the dominant poloidal and toroidal fields. From these, we further found a Fokker-Planck equation satisfied by the dipole field, the coefficients of which we were able to compare directly with data. We believe that this framework will prove useful in analyzing and characterizing observations, Laboratory experiments, and numerical simulations. There are also many ways in which this work can be extended, such as including more modes beyond just the dipole poloidal and quadrupole toroidal, and considering a more detailed physical model of the core.

The work of CRS was performed under the auspices of the U.S. Department of Energy at the Lawrence Livermore National Laboratory under Contract No. DE-AC52-07NA27344. BB acknowledges the support of National Science Foundation grant EAR-1644644.

\bibliography{Scullard2017.bib}
\end{document}